\documentstyle[aps,12pt]{revtex}

\begin{document}
\author{O. B. Zaslavskii}
\address{Department of Physics, Kharkov State University, Svobody Sq. 4, Kharkov\\
310077\\
Ukraine\\
PACS number(s): 04.60.Kz, 04.70.Dy, 04.62.+v}
\author{e-mail: olegzasl@aptm.kharkov.ua}
\title{Two-dimensional self-consistent quantum-corrected geometries with a constant
dilaton field}
\maketitle

\begin{abstract}
It is argued that the existence of constant dilaton field solutions is a
generic feature of string-inspired dilaton gravity. Such solutins arise in
the extreme limit of black hole metrics. It is shown that in a strong
coupling region quantum effects give rise to two horizons in thermal
equilibrium that has no classical counterpart.
\end{abstract}

The existence of black hole solutions in two-dimensional (2D) dilaton
gravity \cite{2d} gives us hope for better understanding some key issues of
black hole physics (singularities, information loss, etc.) which are
difficult to analyze in the more realistic 4D case. One of such issues is
the nature of the extreme state with a zero surface gravity which is
believed to represent an object qualitatively different from nonextreme
black holes either in topological or thermodynamic respects \cite{ross}.
Quantum-corrected geometry of 2D extreme black holes was analyzed in \cite
{trivedi}. Meanwhile, one important class of solutions with a zero surface
gravity was overlooked in that study - solutions with a constant dilaton
field. The existence of such kind of solutions was pointed out in \cite
{frolov} but for the classical case only. In the quantum case the similar
solution was found \cite{solod} for the RST model \cite{rst}.

The aim of the present paper is to show that the existence of solutions with
a constant dilaton field is a generic feature of 2D dilaton gravity.
(Although I deal with the standard form of the action generalization is
straightforward.) Solutions under discussion are of special interest since
they describe the metric near a horizon of extreme 2D black holes (as
indicated in \cite{frolov} with the reference to R. B. Mann's observation).
In this sense the relationship between extreme 2D black holes and solutions
in question resemble that between the Bertotti-Robinson spacetime and
extreme Reissner-Nordstr\"{o}m 4D black holes (see, for example, \cite{and}
and references therein) that may promote a better understanding of what
happens to a 4D black hole dressed by quantum fields near the extreme state.
However, in contrast to the 4D case where the computation of the
stress-energy tensor is a very complicated problem, in the 2D spacetimes it
is known exactly. Moreover, it turns out that, as we will see below, field
equations with backreaction taken into account are also solved for a
constant dilaton field exactly to give a very simple result. It is worth
stressing that, whereas the RST model for which solutions at hand were found
in \cite{solod} is exactly solvable itself, we need not in what follows
imposing such a restriction: in general filed equations with inhomogeneous
dilaton are not solvable but, nevertheless, ''degenerate'' solutions under
discussion can be found exactly.

In classical 2D dilaton gravity they may have one of the following forms 
\cite{frolov}: 
\begin{equation}
ds^{2}=-dt^{2}\exp (-2\alpha l)+dl^{2}  \label{1}
\end{equation}
\begin{equation}
ds^{2}=-dt^{2}\sinh ^{2}\alpha l+dl^{2}  \label{2}
\end{equation}
More exactly, there exist also solutions with $\cosh ^{2}al$ at $dt^{2}$
but, as there is no horizon in that case we will not discuss them further.
Here $\alpha =\sqrt{\lambda }$ where $\lambda >0$ is a cosmological
constant. The metric \ref{1} with the zero surface gravity corresponds to
extreme black holes in the sense explained above. It is interesting that the
metric \ref{2} is intimately connected with the issue of extreme black holes
too: as follows from results of \cite{zasl97}, such a form of the metric
arises in the extreme limit of nonextreme black holes. Thus, both metrics
are related to the issue of the extreme state describing situation in both
topological sectors. Below we will see that quantum effects (which are not
supposed to be small) leave the qualitative character of \ref{1}, \ref{2}
unchanged but give rise to one more type of solutions which was absent in a
classical theory.

Let a system be described by the action $I=I_{0}+I_{q}$ where $I_{0}$ is the
standard action of 2D dilaton gravity and $I_{q}$ is a Polyakov-Liouville
action \cite{pol}. According to \cite{2d} 
\begin{eqnarray}
I_{0} &=&\frac{1}{2\pi }\int d^{2}x\sqrt{-g}e^{\Phi }(R+L)  \label{action} \\
L &=&(\nabla \Phi )^{2}+\lambda -\frac{1}{2}F_{\mu \nu }F^{\mu \nu } 
\nonumber
\end{eqnarray}
Here $F_{\mu \nu }$ is electromagnetic field. We do not write down
explicitly the boundary terms which are necessary for the variation
procedure to be self-consistent. In two dimensions $F_{\mu \nu }=Fe_{\mu \nu
}$ where $e_{\mu \nu }=e_{\left[ \mu \nu \right] },$ $e_{01}=\sqrt{-g}.$ The
electromagnetic-field equation reads 
\begin{equation}
(e^{\Phi }F)_{,\alpha }=0  \label{electro}
\end{equation}
The field equation which is obtained by varying $\Phi $ has the form 
\begin{equation}
2\Box \Phi +(\nabla \Phi )^{2}-R-\lambda -F^{2}=0  \label{dilaton}
\end{equation}
Varying a metric we obtain field equations $T_{\mu \nu }=0$ where $T_{\mu
\nu }\equiv 2\frac{\delta I}{\delta g^{\mu \nu }}.$ Explicit calculation
gives for the classical part of $T_{\mu \nu }:$ 
\begin{eqnarray}
T_{1cl}^{1} &=&\frac{e^{\Phi }}{\pi }(F^{2}+U-\Phi _{;1}^{;1})  \label{11} \\
T_{0cl}^{0} &=&\frac{e^{\Phi }}{\pi }(F^{2}+U-\Phi _{;0}^{;0})  \label{00}
\end{eqnarray}
Here 
\begin{equation}
U=\Box \Phi +\frac{(\nabla \Phi )^{2}}{2}-\frac{\lambda }{2}-\frac{F^{2}}{2}
\label{u}
\end{equation}
semicolon denotes covariant derivative. It follows from \ref{dilaton} that $%
U=R/2.$

We will look for static solution, so the metric can be represented in the
form 
\begin{equation}
ds^{2}=-dt^{2}\mu ^{2}+dl^{2}  \label{qmetric}
\end{equation}
where $\mu =\mu (l).$ Then the Riemann curvature $R=-2\mu ^{\prime \prime
}/\mu .$ For solutions with a constant $\Phi $ 
\begin{equation}
T_{1cl}^{1}=\pi ^{-1}e^{\Phi }(F^{2}-\mu ^{\prime \prime }/\mu )=T_{0cl}^{0}.
\label{classic}
\end{equation}
Now let us make use the explicit expressions for the quantum part of $T_{\mu
}^{\nu }$ in a static geometry which can be obtained from the
Polyakov-Liouville action \cite{pol}: 
\begin{eqnarray}
T_{1q}^{1} &=&\frac{\kappa }{\pi }(\frac{\mu ^{\prime 2}}{\mu ^{2}}-\frac{%
4\pi ^{2}T_{H}^{2}}{\mu ^{2}}),  \label{stress} \\
T_{0q}^{0} &=&-T_{1q}^{1}-\frac{\kappa R}{\pi }  \nonumber
\end{eqnarray}
where the quantum parameter $\kappa =N/24,$ $N$ is a number of scalar fields
in a multiplet.

Consider first solutions with a zero surface gravity (prototype of extreme
black holes). Then, making substitution $\mu =\exp (\sigma )$ one can infer
form eqs. \ref{classic}, \ref{stress} and $T_{1}^{1}-T_{0}^{0}=0$ the
equation $\sigma ^{\prime \prime }=0$ whence $\sigma =cl$ $+d$, where $c$ is
a constant to be determined, without loss of generality one can put $d=0$ by
a proper rescaling time. Then according to \ref{classic}, \ref{stress} $%
T_{1cl.}^{1}=\pi ^{-1}e^{\Phi }(F^{2}-c^{2}),$ $T_{1q}^{1}=\pi ^{-1}\kappa
c^{2}.$ Taking into account eq. $T_{1}^{1}=0$ and eq. \ref{dilaton} we
obtain 
\begin{eqnarray}
2c^{2}-\lambda -F^{2} &=&0  \label{c} \\
F^{2}-c^{2}+\tilde{\kappa}c^{2} &=&0  \nonumber
\end{eqnarray}
where $\tilde{\kappa}=\kappa e^{-\Phi }.$ Solving \ref{c} one easily obtains 
\begin{eqnarray}
c^{2} &=&\lambda (1+\tilde{\kappa})^{-1}  \label{result} \\
F^{2} &=&\lambda (1-\tilde{\kappa})(1+\tilde{\kappa})^{-1}  \nonumber \\
R &=&-2\lambda (1+\tilde{\kappa})^{-1}  \nonumber
\end{eqnarray}
It is seen from \ref{result} that solutions under considerations exist only
for $\lambda >0$ as it takes place in the classical case \cite{frolov}.
Meanwhile, a qualitatively new phenomenon occurs in a quantum domain when $%
\tilde{\kappa}=1$: $F^{2}=0$ whereas for a classical system cancellation of
electromagnetic field with a constant dilaton one implies that $\lambda =0,$
so spacetime is flat. Thus, in a strong coupling regime effects of back
reaction make it possible the existence of the anti-de Sitter like solutions
without electromagnetic field. If $\tilde{\kappa}>1$ solutions under
consideration are absent.

Let now the temperature be finite. It is convenient to normalize time in
such a way that $T_{H}=(2\pi )^{-1}.$ Repeating calculations step by step,
we obtain that formulae \ref{result} for curvature and electromagnetic field
hold true, the function $\mu =c^{-1}\sinh cl.$ Thus, in both considered
cases quantum corrected geometry reads 
\begin{equation}
ds^{2}=-e^{-2cl}dt^{2}+dl^{2}  \label{q1}
\end{equation}
(quantum analog of \ref{1}) or 
\begin{equation}
ds^{2}=\frac{-\sinh ^{2}cl}{c^{2}}dt^{2}+dl^{2}  \label{q2}
\end{equation}
(quantum analog of \ref{2}) . It is seen that quantum corrections do not
change qualitatively the character of spacetime. In particular, the limiting
transition of a black hole to the extreme state with a temperature finite in
any point outside a horizon is performed in the same manner as for a bare
hole. It is worth stressing that, while \ref{q1} or its classical
counterpart \ref{1} represent the metric of an extreme black hole only in
the vicinity of a horizon \cite{frolov} eq. \ref{q2} describes the metric of
a nearly extreme black hole in the whole manifold as it follows from \cite
{zasl97}.

The most interesting result consists in the possibility to have $\mu
=c^{-1}\sin cl.$ In this case 
\begin{equation}
ds^{2}=-\frac{\sin ^{2}cl}{c^{2}}dt^{2}+dl^{2}  \label{sin}
\end{equation}
\begin{eqnarray*}
c^{2} &=&-\lambda (1+\tilde{\kappa}) \\
R &=&-2\lambda (1+\tilde{\kappa})^{-1} \\
F^{2} &=&-\lambda (\tilde{\kappa}-1)(\tilde{\kappa}+1)^{-1}
\end{eqnarray*}
These solutions exist only if $\lambda <0,$ $\tilde{\kappa}\geq 1$ and have
no classical analog. From the physical viewpoint they represent two horizons
(at $l=0$ and $l=\pi /c)$ in thermal equilibrium. It was shown in \cite
{horizon} that such a metric always arises when radii of a black hole and
cosmological horizons coalesce (but the proper distance between them remains
finite as seen from \ref{sin}) that generalizes a series of observations
made in some particular cases \cite{part}. Now the curvature $R>0$ (de
Sitter type case). Thus, quantum effects change the structure of spacetime
drastically. Although the solution at hand can exists only in a strong
coupling regime and in this sense is pure quantum, the curvature of
spacetime is classical in that it is proportional to the cosmological
constant similarly to what takes place with constant dilaton field solutions
in the RST model \cite{solod}. The quantum coupling parameter in our problem
can take arbitrary values, the electromagnetic field strength being adjusted
to it according to \ref{result} or \ref{sin} whereas in the limit $F=0$ (as
was assumed in \cite{solod}), solutions under discussion exist only for one
preferable value $\tilde{\kappa}=1$ in agreement with \cite{solod} as it
follows from \ref{result}, \ref{sin}. While for solutions \ref{q1}, \ref{q2}
the energy density of quantum fields $-\pi ^{-1}\kappa c^{2}$ is negative
for the spacetime \ref{sin} it is positive and is equal to $\pi ^{-1}\kappa
c^{2}.$

It is worth dwelling upon the following subtlety. When the dilaton is
constant, the curvature term in Lagrangian becomes dynamically irrelevant
because of topological invariance of the Ricci scalar in two dimensions, so
one can wonder where not-flat solutions originate from. It is instructive in
this respect to compare the classical and quantum solutions. In the first
case the nontrivial contribution to field equations comes from the $\lambda $
- term and electromagnetic field, so it is the electromagnetic force only
which curves the geometry. As a result, $R=-2F^{2}$\cite{frolov} as seen
from \ref{result}, so in the limit $F=0$ the curvature goes to zero as well
and the solution is necessarily flat. In the quantum case, however, account
for the Polyakov - Liouville action brings about the new contribution to
field equations: back reaction of quantum fields. As a result, the curvature
is due to either the electromagnetic force or back reaction.

It is worthwhile to note that for all values of the coupling parameter $%
\tilde{\kappa}\neq 1$ the regions with $\tilde{\kappa}<1$ and $\tilde{\kappa}%
>1$ are strongly separated: the metric under discussion can be only of the
anti-de Sitter type in the first case and only of the de Sitter one in the
second case. However, the point $\tilde{\kappa}=1$ is exceptional in the
sense that either the de Sitter or anti-de Sitter solution may exist
depending on the sign of $\lambda .$ As in that case $F=0$, the nonzero
curvature arises due to the back reaction of quantum fields only.

All metrics considered above arise in a self-consistent way, with one-loop
effects of back reaction of quantum fields taken into account exactly. They
possess the following attractive feature: whereas on a horizon of a generic
two-dimensional black hole weak divergencies of stress-energy tensors
inevitably occur \cite{trivedi}, \cite{and2} such divergencies are absent
for spacetimes at hand. Indeed, the relevant quantity $\mu
^{-2}(T_{1q}^{1}-T_{0q}^{0})$ \cite{and2} vanishes for all three solutions 
\ref{q1}, \ref{q2}, \ref{sin}. It gives some hint that such degenerate
solutions could be suitable candidates to the final state of a black hole
after evaporation of it. The considered examples can also serve as a toy
model for a more complicated question about properties of four-dimensional
extreme and nearly extreme black holes dressed by one-loop effects of
quantum backreaction.

\end{document}